\documentclass[aps, prx, showpacs, superscriptaddress, floatfix]{revtex4-1}

\usepackage{graphicx}  
\usepackage{epstopdf}
\usepackage{dcolumn}   
\usepackage{bm}        
\usepackage{amsmath}
\usepackage{amssymb}   
\usepackage{wrapfig}
\usepackage{array}
\usepackage[caption=false]{subfig}
\usepackage[bookmarks=false,linkcolor=blue,urlcolor=blue,colorlinks,citecolor=blue]{hyperref}
\usepackage{exscale,relsize}
\usepackage{color}
\usepackage{times, verbatim}
\usepackage{pstricks}

\usepackage{pst-all}

\begin{document}

\title{Realizing Majorana zero modes in superconductor-semiconductor heterostructures}

\author{R. M. Lutchyn}
\affiliation{Station Q, Microsoft Research, Santa Barbara, California 93106-6105, USA}

\author{E. P. A. M. Bakkers}
\affiliation{QuTech and Kavli Institute of Nanoscience, Delft University of Technology, 2600 GA Delft, The Netherlands}
\affiliation{Department of Applied Physics, Eindhoven University of Technology, 5600 MB Eindhoven,
The Netherlands}

\author{L. P. Kouwenhoven}
\affiliation{QuTech and Kavli Institute of Nanoscience, Delft University of Technology, 2600 GA Delft, The Netherlands}
\affiliation{Microsoft Station Q at Delft University of Technology, 2600GA Delft, The Netherlands}

\author{P. Krogstrup}
\affiliation{Center for Quantum Devices and Station Q Copenhagen, Niels Bohr Institute, University of Copenhagen, Copenhagen, Denmark}

\author{C. M. Marcus}
\affiliation{Center for Quantum Devices and Station Q Copenhagen, Niels Bohr Institute, University of Copenhagen, Copenhagen, Denmark}

\author{Y. Oreg}
\affiliation{Department of Condensed Matter Physics, Weizmann Institute of Science, Rehovot 76100, Israel}

\date{\today}

\begin{abstract}

Realizing topological superconductivity and Majorana zero modes in the laboratory is one of the major goals in condensed matter physics.  We review the current status of this rapidly-developing field, focusing on semiconductor-superconductor proposals for topological superconductivity. Material science progress and robust signatures of Majorana zero modes in recent experiments are discussed. After a brief introduction to the subject, we outline several next-generation experiments probing exotic properties of Majorana zero modes, including fusion rules and non-Abelian exchange statistics. Finally, we discuss prospects for implementing Majorana-based topological quantum computation in these systems.

 \end{abstract}

\maketitle


\section{Introduction}

The search for topological phases of matter has generated significant interest in physics, chemistry, and material science~\cite{Wilczek'09, SternPerspective,
Brouwer_Science, Lee14}. The discovery of topological superconductors supporting Majorana zero-energy modes is of fundamental scientific importance and has profound technological applications for quantum information processing. Majorana zero modes correspond to exotic {\it neutral} excitations of a superconductor comprising of an equal superposition of an electron and a hole. In this sense, a solid-state Majorana quasiparticle which is its own antiparticle is closely related to a Majorana fermion that was first introduced by Ettore Majorana in the context of elementary particle physics~\cite{Majorana2008}. A {\it neutral} excitation in a superconductor has a very special property due to the inherent particle-hole symmetry - it is bound to zero energy so that there is no cost to occupy such a state. Therefore, the presence of a large number of localized Majorana zero-energy modes leads to a massive ground-state degeneracy which can be exploited for topological quantum computing purposes~\cite{kitaev2003fault, Nayak08}. Theory~\cite{Moore1991, ReadGreen} predicts that Majorana zero-energy modes (MZMs) localized at defects obey an exotic exchange statistics, similar to that of non-Abelian anyons. This is by far the most interesting property of MZMs. Indeed, exchanging the position of MZMs corresponds to a nontrivial transformation within the degenerate ground-state manifold, and represents a non-commutative operation which does not depend on the way and the details of its execution. Therefore, such an operation is topologically protected (i.e. only depends on exchange statistics of the quasiparticles) and may be used to implement quantum gates. These ideas are at the heart of the topological approach to quantum computing~\cite{kitaev2003fault, Nayak08}. Indeed, by exploiting topological materials, which by their nature minimize errors~\cite{kitaev2003fault}, one may overcome the largest barrier to building scalable quantum computing - decoherence.


A number of platforms for realizing MZMs in the laboratory were recently put forward~\cite{Qi'08, Beenakker13a, Alicea12a, Leijnse12, Stanescu13b, Franz'15, DasSarma15, MasatoshiSato2016, Aguado2017}. Most of them propose to engineer an appropriate model Hamiltonian at the interface of a conventional superconductor and some other materials such as topological insulators~\cite{FuKane, Fu&Kane09, Cook11, Sun'16}, semiconductors with strong spin-orbit coupling~\cite{Sau2010, Alicea10, Lutchyn10, Oreg10, SukBum'11, Duckheim'11, Potter'12} and magnetic atom chains~\cite{Choy'11,nadj2013, klinovaja2013,braunecker2013,vazifeh2013,pientka2013, Nakosai2013, Kim14, brydon2014, Li14, Kotetes2014, Ojanen2015, nadjperge2014,Ruby15,Pawlak'16, J.Zhang'16}. Among the most promising ones are the proximitized nanowire proposals~\cite{Lutchyn10, Oreg10}, discussed in details in this review. There is mounting experimental evidence for the presence of MZMs in proximitized nanowires~\cite{Mourik12,Rokhinson12,Deng12,Churchill13,Das12,Finck12,Albrecht16,Zhang16,Chen2016, Deng2016, Suominen2017, Nichele2017, Zhang2017, Zhang2017a, Sestoft2017, Deng2017, Geresdi2017}.

A paradigmatic model for a one-dimensional (1D) topological superconductor involves spinless fermions that hop along the chain and experience proximity-induced p-wave pairing~\cite{kitaev01}. Such a system supports an odd number of localized MZMs at the opposite ends of the chain. As explained above, these modes are neutral
and, thus, a single MZM cannot accommodate a conventional (Dirac) fermion. At least two MZMs are needed to form a fermionic state with a well-defined occupation number. As a result, a 1D topological superconductor supports a non-local fermionic mode comprising of two MZMs localized at the opposite ends and separated by a distance which can be much larger than the superconducting coherence length. This non-local entanglement in a gapped system is the hallmark of topological superconductors and represents an instance of electron fractionalization~\cite{kitaev01}. Note that the above argument relies on having an odd number of MZMs per wire end since an even number of Majorana modes can pair up and form a conventional Andreev state locally. In the latter case, the fermionic mode generically resides at a finite energy and, as a result, there is no non-local entanglement in the ground-state. Thus, there is a profound difference between conventional gapped superconductors, which have a unique ground state with even fermion parity, and topological superconductors, which instead have a highly degenerate ground state due to the presence of many MZMs \cite{ReadGreen, kitaev01}.

\begin{figure}[hb]
\includegraphics[width=0.9\columnwidth]{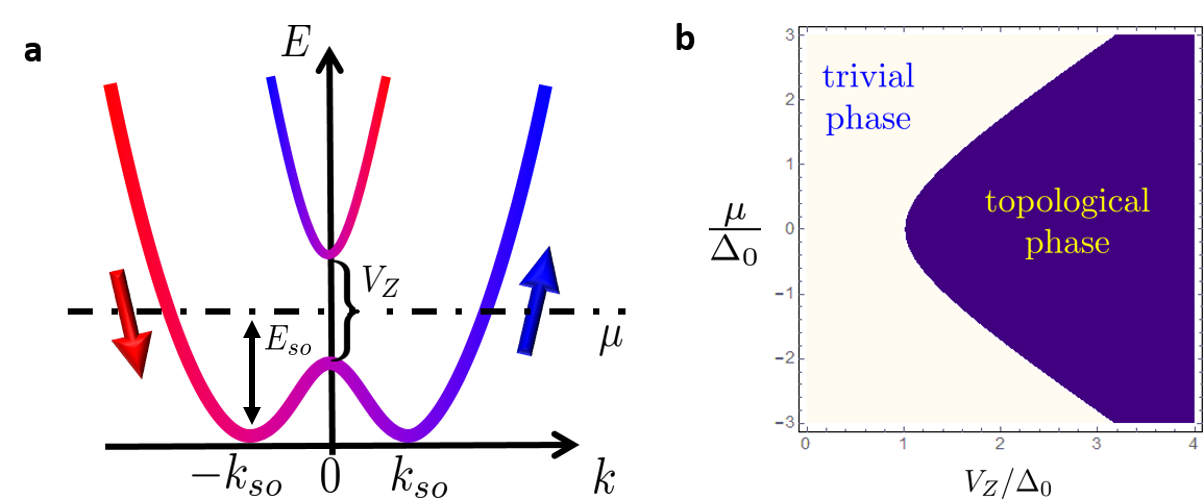}
\caption{a) Energy spectrum as a function of the momentum along the nanowire $k$. Spin-orbit coupling shifts the parabolas describing electron spectrum sideways by $k_{so}$ and introduces new energy scale $E_{so}$. Zeeman coupling $V_Z$ due to applied external magnetic field perpendicular to the direction of the spin-orbit coupling opens a gap at $k\!=\!0$~\cite{Lutchyn10, Oreg10}. Arrows indicate approximate spin orientation for different momenta. b) Topological quantum phase diagram. By changing chemical potential $\mu$ in the nanowire or Zeeman splitting $V_Z$ one can drive the system into a topological phase. Here $\Delta_0$ is the induced pairing potential in the wire.}
\label{fig:fig1a}
\end{figure}

Spinless p-wave superconductivity is a key element for realizing separated MZMs. However, electrons in conventional materials have spin $\frac{1}{2}$, and, thus, the notion of a spinless superconductor does not seem immediately relevant to real physical systems. An elegant way to overcome this difficulty is to use spin-orbit materials, where spin and orbital degrees of freedom are correlated. Indeed, spinless superconductivity can effectively emerge in a semiconductor nanowire with strong spin-orbit interaction proximity-coupled to a conventional (s-wave) superconductor~\cite{Lutchyn10, Oreg10}. The corresponding energy spectrum for the nanowire with spin-orbit coupling is shown in Fig.~\ref{fig:fig1a}a. The spin-orbit interaction shifts the momentum parabolas sideways, the Zeeman term, which does not commute with the spin-orbit term, opens a gap in the spectrum at zero momentum, see Box~\ref{box} for details. When the chemical potential $\mu$ is tuned to be in the gap (see Fig.~\ref{fig:fig1a}a), the spectrum has a single band crossing Fermi points and, thus, the system in this regime is effectively described by the ``spinless'' degrees of freedom. However, due to the spin-orbit-induced rotation of the spins at the opposite Fermi points, see Fig.\ref{fig:fig1a}a, proximity-induced s-wave interaction opens a pairing gap in the spectrum. The resulting state is closely related to a spinless p-wave superconductor as explained in the Box~\ref{box}.

One of the virtues of the above model is that the proximitized nanowire can be driven into a topological phase by tuning magnetic field or chemical potential, see Fig.\ref{fig:fig1a}b. The emergence of MZMs at a certain critical value of a control parameter is necessarily accompanied by the closing of the bulk gap~\cite{ReadGreen, kitaev01}. This phenomenon corresponds to a topological quantum phase transition (i.e., a quantum phase transition between topologically trivial and non-trivial states, see Fig.\ref{fig:fig1a}b). The topological phase is stable with respect to small perturbations (e.g., disorder) as long as these perturbations do not collapse the bulk gap in the spectrum. The latter depends on the effective spin-orbit energy $E_{so}$, the proximity-induced gap $\Delta_0$, and the effective Zeeman energy $V_Z$ in the heterostructure. Therefore, when engineering such materials, an obvious first choice is to pick material components that individually have required properties. So far, the heavy-element semiconductors InAs and InSb have received much attention due to the strong spin-orbit coupling as well as large Lande $g$-factor, whereas Al and NbTiN have been primarily used for the superconducting components. Relevant bulk properties of these materials are listed in Table~\ref{Table1}. The large $g$-factor in the semiconductor allows one to achieve the desired effect, opening a large effective Zeeman gap in the hybrid nanowire using an in-plane magnetic field at fields below the critical field of the s-wave superconductor. High-quality interfaces between the two components lead to a significant hybridization of the semiconducting and metallic states resulting in a large induced pairing potential. The interface should be smooth since disorder scattering tends to suppress p-wave superconductivity and eventually leads to the collapse of the topological phase~\cite{Motrunich01, Brouwer11, Stanescu11, Akhmerov2011, Potter12, Lobos12, Potter_Lee2011, lutchyn_momentum, DeGottardi2013, Takei13, Adagideli14, Hui15, Cole16, Hegde2016, DongLiu2017}. All these requirements are important for the observation of MZMs and pose a challenging material science task.

\begin{table}
\centering
\begin{tabular}{|c|c|c|}
  \hline
  Semiconductors & InAs & InSb  \\
  \hline
  $g$-factor & 8-15 & 40-50 \\
  effective mass $m^*$ & 0.023 $m_e$ & 0.014 $m_e$ \\
  spin-orbit energy $E_{so}=\frac{m^* \alpha^2}{2\hbar^2}$ & 0.05 - 1 meV & 0.05 - 1 meV \\
  spin-orbit coupling $\alpha$ \, & 0.2 - 0.8 eV$\cdot$ {\AA}  & 0.2 - 1 eV$\cdot$ {\AA} \\
  spin-orbit length $\lambda_{so}\!\equiv\! k^{-1}_{so}\!=\!\frac{\hbar^2}{\alpha m^*}$ & 180 - 40 nm & 230 - 50 nm \\
  \hline
  Superconductors & Al & NbTiN \\
  \hline
  superconducting gap  $\Delta$ & 0.2 meV & 3 meV \\
  \ critical field $B_c$ & 10 mT & 10 T \\
  \ critical temperature $T_c$ & 1.2 K & 15 K \\
  \hline
\end{tabular}
\caption{Bulk properties of typically used semiconductors and superconductors~\cite{levinshteinhandbook, Winkler2003, Cochran1958, Mourik12}. Rashba spin-orbit coupling strength was measured in nanoscale structures in Refs.~\cite{Shabani16, Weperen2015}.}
\label{Table1}
\end{table}

In this article, we provide a perspective on the current state of the realization of MZMs in superconductor (SC)-semiconductor (SM) heterostructures. In the next section, we will provide an overview of the recent material science developments in growing high-mobility semiconductors and preparing high-quality interfaces between superconductors and semiconductors. We then summarize robust signatures of MZMs and discuss the best experimental evidence for the presence of MZMs in proximitized nanowires. Finally, we conclude with an outlook and discuss the perspective for topological quantum computation with MZMs. Due to space constraints, we will not cover topological insulator and magnetic-atom-chain Majorana proposals. There are several excellent reviews on this subject~\cite{Qi'08, Hasan'10, Murakami'11,Ando'13, bernevig2013topological, Bansil'16} to which we refer an interested reader.

\section{Material science aspects}

\subsection{Semiconductor growth}

\begin{figure}
\includegraphics[width=\columnwidth]{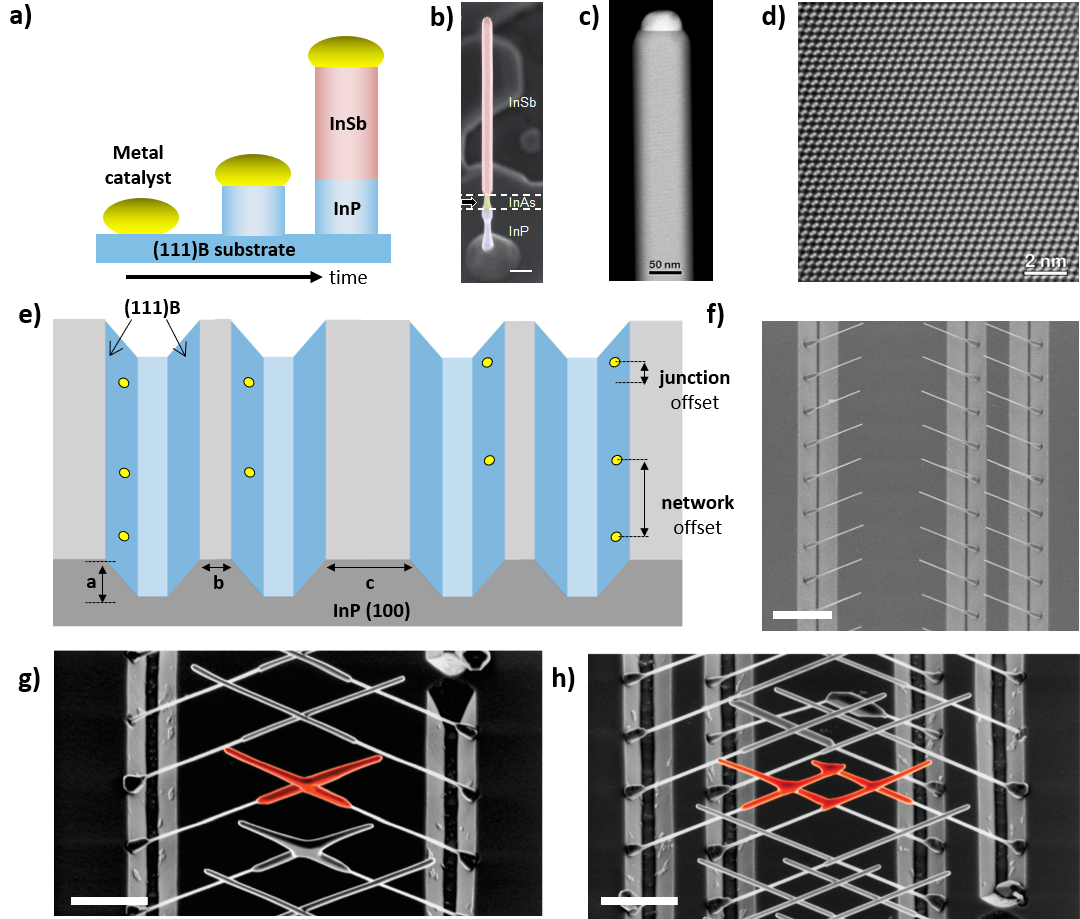}
\caption{Growth of bottom-up nanowires. a) Vapor-Liquid-Solid (VLS) nanowire growth mechanism, b) SEM image of an InP-InSb nanowire grown along the (111)B crystal direction, c) TEM image showing the defect-free nature of the InSb wire, d) HRTEM image demonstrating the (111) growth direction, e) generic method to fabricate nanowire network with a pre-defined number of wires. Trenches are etched from a (100)-oriented InP substrate such that (111)B facets are exposed. E-beam lithography is used to define catalyst on the (111)B facets with a controlled position. SEM images of f) InP nanowire grown from the catalysts on the sloped facets, g) InP-InSb nanowires forming a cross, h) InSb nanowire hashtags. Scale bars are $1 \mu m$. Panels e)-h) reproduced with permission from Ref.~\cite{Gazibegovic2017}.}
\label{fig:fig2}
\end{figure}

The fabrication of one-dimensional confinement in the InAs and InSb semiconducting materials has been obtained by two different methods. One is top-down (lithography-based) definition of 1D structures in two-dimensional (2D) quantum wells grown by Molecular Beam Epitaxy (MBE). Here the quantum wells are grown close to the surface allowing for the deposition or epitaxial growth of a superconducting material in close proximity to the well (see section \ref{episuper}). The growth of 2D semiconductor materials has been studied for many years and is now well understood for many different material combinations. However, for SM-SC heterostructures new challenges emerge. To increase the mobility in conventional 2D materials, the first thing would be to grow a thick insulating, and, ideally, lattice-matched layer on top of the 2DEG to protect it from the surface disorder. However, this will also suppress induced superconductivity which relies on electron exchange between SM and SC. Thus, a thinner and low barrier material is grown as an intermediate layer, and the task is to find the barrier height and thickness that give the best combination of mobility and induced superconductivity~\cite{Shabani16}. Moreover, as the 1D confinement will be defined by top-down methods using either mesa etch or selective etch of the superconductor, there will be additional material/etching requirements for allowing a controlled definition of the nanowires. This is likely to be an ongoing research topic for the coming years.

The other way to produce nanowires is to use the bottom-up Vapor-Liquid-Solid (VLS) growth mechanism.
This mechanism has been developed over the last 50 years and can be used to fabricate nanowires of a variety of semiconductor materials~\cite{Wagner64, Caroff09, Lugani10, Nilsson10, Vogel11, Plissard12}. InAs and InSb materials have also shown promising behavior in this fabrication scheme. Typically, a catalyst particle is used to collect precursor material and to establish a local supersaturation, as shown in Fig.~\ref{fig:fig2}a. The catalyst becomes liquid, and when a critical supersaturation level is reached, the material is precipitated at the liquid/substrate interface by a layer-by-layer growth mechanism. This mechanism results in a crystalline nanowire with the radius determined by the catalyst dimensions. Precursor material can directly impinge from the gas phase onto the catalyst surface, or it can first land on the substrate surface and subsequently diffuse towards the catalyst. While diffusing, these adatoms can be trapped, and when this happens on the side wall of the nanowire it will lead to radial growth. Since we want to fabricate 1D nanowires, it is important to suppress this effect. In the case of InSb nanowire growth, a problem is the low surface energy of Sb, which acts as a surfactant and, therefore, makes it difficult to initiate nanowire growth because the wetting layer tends to form a 2D layer. One way to overcome these problems is by growing a heterostructure. For example, as shown in Fig.~\ref{fig:fig2}b, an InP/InAs stem wire can be used to grow an InSb wire~\cite{Caroff09, Lugani10, Nilsson10, Vogel11, Plissard12}. These wires can be a few microns long and have a diameter of 80-100 nm.

The crystal structures of VLS grown nanowires are relatively easily investigated by transmission electron microscopy (TEM). Fig.~\ref{fig:fig2}d shows a defect-free cubic (zincblende) structure of a InSb nanowire grown in the [111]B direction. On the other hand,  defect-free InAs nanowires are typically grown in a hexagonal (wurtzite) crystal phase with a [001]B axis. It is important to mention that nanowires have a strong tendency to grow in [111]B/[001]B direction, which can be used to direct the orientation of the wires as discussed below. These uniform wires have been used for the first-generation Majorana experiments~\cite{Mourik12}, and their electronic properties are discussed in Fig.~\ref{fig:fig4}. For next-generation Majorana experiments more complex nanowire structures are required, for instance, T-junctions as explained in Sec.~\ref{sec:perspective}. An important requirement is that the junction between the branches has small reflection amplitude for an incoming electron. While such structures can be processed ex-situ using top-down lithography in 2D materials, VLS grown structures can be fabricated in-situ by using the crystallography of the substrate~\cite{plissard13,Kang13, Car14, Dalacu13, Heedt16, Rieger16,Krizek17}. For example, instead of a (111) substrate we can use a (001) substrate where the growth direction can be turned into a non-vertical direction. Instead of a (111) substrate one can, for example, use a (001) substrate, in which trenches are etched exposing the (111)B facets. By using e-beam lithography the catalysts are defined on the sloped facets as depicted in Fig.~\ref{fig:fig2}e. Nanowires will grow perpendicular from these facets into two equivalent $< 111 >$B directions as shown in Fig.~\ref{fig:fig2}f. After growth of the InP stem, the InSb segment is grown. Since all wires are epitaxially connected to the substrate, they will, by definition, form a single crystal when
they merge. For the merging process, there are two important aspects to be considered. First, there should be a junction offset in catalyst position as indicated in Fig.~\ref{fig:fig2}e. If the offset is too small, the catalyst particles will merge together, and a nanowire bridge is formed; the large Au particle crawls on a nanowire side facet. If the offset is too large the wires will miss each other. An optimum offset equals the wire radius. Second, the merging process itself is governed by radial growth, which, as mentioned above, always takes place as a competitive reaction besides axial growth~\cite{Car14}. This radial growth results in an epitaxial shell and is responsible for the merging of two nearby wires. In the SEM images in Fig.~\ref{fig:fig2}g and h it is clear how wires can merge to form crosses or hashtags. With this method, a predefined number of wires can be connected into a network, and the dimensions and symmetry of the resulting structures can be tuned by the spacings b and c as indicated in Fig.~\ref{fig:fig2}e. TEM studies reveal the single crystalline nature of the crossed wires.

\subsection{Superconducting interfaces}\label{episuper}

\begin{figure}
\center
\includegraphics[width=\columnwidth]{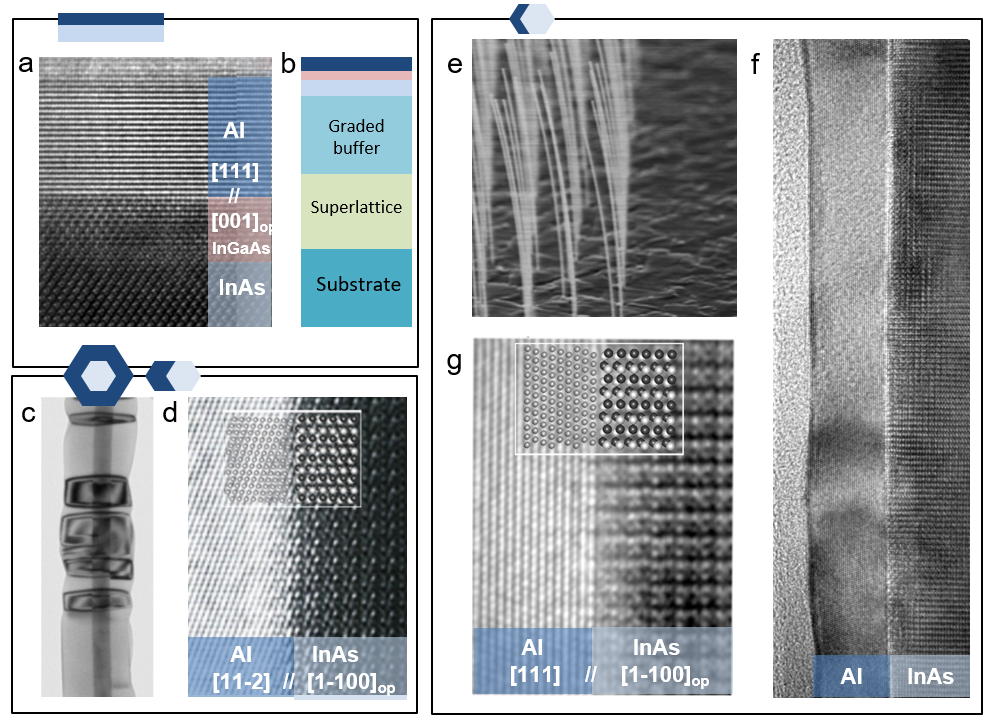}
\caption{High symmetry epitaxial Al/InAs interfaces. a) For planar InAs-based materials, the dominating Al fcc crystal orientation at low temperatures is [111] out of plane. Unlike the nanowire crystals, the planar epitaxial heterostructure shows better performance with an InGaAs barrier layer between the Al and InAs. b) A typical example of a layer stack grown on a lattice mismatched substrate like InP is shown. Panels a)-b) adapted with permission from Ref.~\cite{Shabani16}. Here the superlattice traps the impurities and the graded buffer makes the transition to the lattice constant of the active region c) Thick Al layers on the nanowire facets typically lead to a gradual transition to a [11-2] out-of-plane crystal reconstruction. The image shows dark and white contrasts which denote two different degenerate twin orientations. The resulting Al can form a complete coherent InAs/Al bicrystal phase with a near lattice matched 3:2 ratio as shown in d). e) SEM micrograph of an array of InAs/Al nanowires where the Al sits on two facets. The nanowires bend towards the Al covered side. f) A thin uniform Al phase is obtained with [111] Al crystal orientation out-of-plane, as seen in the high-resolution zoom on the interface in g). Panels c)-g) adapted with permission from Ref.~\cite{Krogstrup15}. }
\label{fig:fig3}
\end{figure}

First-generation Majorana experiments ~\cite{Mourik12, Deng12, Churchill13, Das12, Finck12} showed a large subgap conductance indicating unintended subgap states, which are detrimental to topological quantum computation as they degrade topological protection. A likely source of the so-called soft gap, or continuum of subgap states, is disorder at the semiconductor/superconductor interface~\cite{Takei13}. For these devices, the native oxide forming when the wires are exposed to ambient air has to be removed prior to the deposition of the superconductor material. There has been significant progress in improving these interfaces by gently etching the oxide and optimizing the superconducting deposition~\cite{Krogstrup15, Zhang16, Zhang2017a}. An important step forward was taken with epitaxial growth of thin aluminum films on pristine facets of the nanowires, without breaking vacuum~\cite{Krogstrup15}. As shown in Fig. 4c, this method of in-situ superconductor growth yields further significant suppression of the subgap density of states~\cite{Chang14}. Subsequently, high-quality epitaxial superconductor-semiconductor interfaces have been realized in other heterostructures~\cite{Shabani16, Gazibegovic2017, Beidenkopf2017, Gusken2017, Sestoft2017}. Typical values of the proximity-induced superconducting gap in different semiconductor-superconductor heterostructures are listed in Table~\ref{Table2}.

The basic requirements for the Al film, besides the ability to induce superconductivity, are, first, it should stay thin enough to withstand high parallel magnetic fields without being driven normal, and second, it should have a uniform morphology along the nanowire, ensuring translational invariance in order to avoid disorder-induced subgap states. In-situ shadow deposition can be used to define junctions in the superconductor without additional lithography and etching steps, avoiding detrimental disorder from top-down processing~\cite{Gazibegovic2017,Krizek17}.


\begin{table}[h]
\centering
\begin{tabular}{|c|c|c|c|}
  \hline
   semiconductor-superconductor heterostructure & InSb-NbTiN & InSb-Al & InAs-Al  \\
  $\Delta_0$, meV & 3 & 0.2-0.3 & 0.2-0.3 \\
  max[$\Delta_{\rm ind}$], meV & 1 & 0.2 & 0.2 \\
  \hline
\end{tabular}
\caption{Typical values for the bulk $\Delta_0$ and maximum values for the proximity-induced $\Delta_{\rm ind}$ superconducting gaps in different semiconductor-superconductor heterostructures~\cite{Zhang16, Deng2016, Zhang2017}. SC gap of Al depends on the film thickness~\cite{Court2008}. Note that $\Delta_{\rm ind}$ in aluminum heterostructures is close to the bulk superconducting gap indicating on strong tunneling regime between semiconductor and superconductor.}
\label{Table2}
\end{table}

The equilibrium shape of Al on a large rigid InAs surface is not a continuous film, but a dewetted droplet with a contact angle of about 60-80 degrees. However, if the characteristic migration length of the Al adatoms is much shorter than the thickness of the nominally deposited Al, it can be kinetically locked to a continuous film. The key to obtaining a short adatom migration length is lowering the temperature sufficiently below room temperature. Moreover, because the native oxide stabilizes the film morphology, it is important to take the sample out of vacuum before it reaches dewetting temperatures (the specific transition temperature depends on the type of interface), which for Al on InAs is typically below room temperature. Unlike the InAs growth, where the crystal structure and morphology are locked once it is formed, the Al phase can easily recrystallize through grain growth transitions, which, at the right temperatures, moves the grain boundaries while keeping film continuous.

The grain growth process is driven exponentially by thermodynamic excesses such as interfacial and strain-free energies. The measure of the thermodynamic driving force is the excess chemical potential, which for a given grain M in the thin film limit can be divided into four terms corresponding to the surface, SC-SM interface, grain boundary (GB) and strain contributions~\cite{Krogstrup15, Shabani16}:
\begin{equation}\label{eq:1}
\mu_M\propto \frac{\gamma_{\rm surface}}{h}+\frac{\gamma_{\rm SC-SM}}{h}+\frac{\gamma_{\rm GB}}{R}+\frac{S \varepsilon^2 }{1-\nu}.
\end{equation}
Here $\gamma_i$ is the energy density of the $i$’th interface, $h$ is the film thickness, $S$, $\varepsilon$ and $\nu$ are the shear modulus, strain and Poisson ratio of the thin film material, respectively; $R$ is the mean in-plane radius of curvature of $M$.

The epitaxial growth of Al on the facets of free-standing InAs nanowires is a more complex and dynamical process than on a planar InAs substrate, as was described subsequently for a two-dimensional platform in Ref.~\cite{Shabani16}.
In the single interface case, the dominating term for thin uniform layers (when $h \ll R$ ) is the surface term, because $\gamma_{\rm surface} > \gamma_{\rm SC-SM}$, which fixes the out-of-plane orientation to the lowest energy facet, which is the (111) facet, see Fig.~\ref{fig:fig3}a. The in-plane orientation appears to be driven primarily by the SC-SM interfacial bi-crystal symmetries which are governed by the second term. As shown in Ref.~\cite{Shabani16}, only two degenerate in-plane grain orientations of Al on planar InAs (100) were observed, indicating that these orientations are local minima of the interface energy with the (111) out-of-plane constraint.
The size of the grains, however, is driven by minimizing the grain boundary energies, governed by the third term in Eq.~\eqref{eq:1}. The last term in Eq.~\eqref{eq:1} could play an important role if a particular bicrystal lattice match at the SC-SM interface has a substantial residual lattice mismatch but a significant minimum in the respective interfacial energy for the particular match. However, for a critical value of the film thickness $h \approx 5$nm Al phase on a planar InAs substrate was measured to be fully relaxed~\cite{Shabani16}, which means that the critical thickness for relaxation is presumably much less, and therefore does not play an important role in the growth kinetics for this particular material combination.
For free-standing nanowires with Al growing on multiple facets, the grain boundary energies play a dominant role in the overall orientation, especially at thicker phases. As the film gets thicker the Al recrystallizes from a (111) out-of-plane orientation, similar as in the case of a single plane, to an orientation that lowers the total amount of grain boundary energy. In other words, the Al orientation changes so that it matches the symmetry of the nanowires and minimizes the incoherent grain boundary excesses across the adjacent facets, see Fig.~\ref{fig:fig3}c and d. The surface of the Al gets more uneven because it still has a tendency to form non-planar (111) facets. However, thin Al, as needed to withstand high magnetic fields, can stay extremely flat and uniform, see, for example, Fig.~\ref{fig:fig3}e-g.

\subsection{Characterization of superconductor-semiconductor heterostructures}

\begin{figure}
\center
\includegraphics[width=\columnwidth]{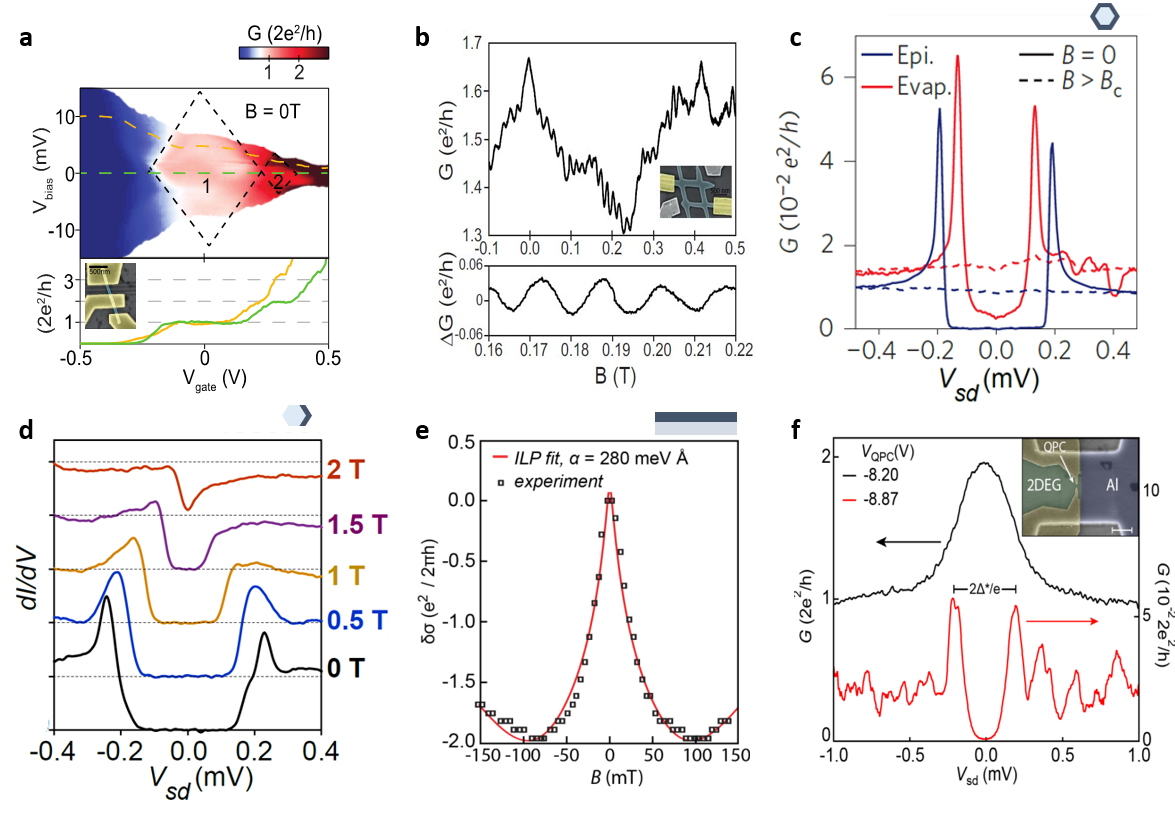}
\caption{Characterization of different materials. a) Quantized conductance through an InSb wire at zero magnetic field. Inset: SEM image of an InSb nanowire with ohmic contacts. Panel a) adapted with permission from Ref.~\cite{Kammhuber16}. b) Aharonov-Bohm oscillations in a nanowire hashtag demonstrating phase coherent transport.  Inset: SEM image of an InSb nanowire hashtag. Panel b) adapted with permission from Ref.~\cite{Gazibegovic2017}. c) Differential conductance as a function of the source-drain voltage of an epitaxial full-shell InAs/Al nanowire (blue) compared with an ex-situ evaporated Al shell (red) at $B = 0$ (solid line) and above the critical field $B > B_c$ (dashed line). Panel c) adapted with permission from Ref.~\cite{Chang14}. d) Tunneling spectra show hard induced gap beyond 1.5~T in an InAs nanowire with a two-facet epitaxial Al film. Panel d) adapted with permission from Ref.~\cite{Deng2016}. The wire is depleted by a gate and shows no subgap states in this regime. e) Weak antilocalization measurements on a hall bar with a 5nm InGaAs barrier and Al selectively etched away. The data allows one to extract the magnitude of the SOC in the semiconductor. Panel e) adapted with permission from Ref.~\cite{Shabani16}. f) Tunneling conductance in the high density (black line) and low density (red line) regimes. The latter provides an estimate for the induced superconducting gap. Panel f) adapted with permission from Ref.~\cite{Kjaergaard2016}. }
\label{fig:fig4}
\end{figure}

Together with strong spin-orbit coupling and induced superconductivity, a key requirement for MZMs is quasi-ballistic electron transport with a controlled odd number of occupied modes. In the ballistic regime, the motion of 1D confined electrons is restricted to discrete energy bands resulting in quantized conductance plateaus. The observation of quantized conductance in the nanowires, therefore, provides direct evidence for the quasi-ballistic transport in these nanowires and allows one to control the mode occupation. Quantized conductance is difficult to achieve in nanowires because impurities, as well as scattering at the nanowire/contact interface, result in an increased probability of reflection in quasi-1D geometry, obscuring the observation of quantized conductance. In order to have highly transparent contacts, InSb wires have been exposed to a sulfur etch to remove the oxide layer before Cr/Au contact deposition. The resulting nanowire device, shown in the inset of Fig.~\ref{fig:fig4}a, consists of bottom gates on which an h-BN flake is deposited as a high-quality dielectric~\cite{Kammhuber16}. The samples are mounted in a dilution refrigerator with a base temperature of 15 mK. The differential conductance $G$ is measured as a function of gate voltage and bias voltage, see Fig.~\ref{fig:fig4}a. At zero bias voltage, an extended plateau is visible at $ G_0=2 e^2/h$, demonstrating ballistic transport through the lowest nanowire subband. From these measurements, one can extract a mean-free-path $l \approx 300$nm. An additional small plateau at $2G_0$ is observed, indicating that the device has a small, but finite energy splitting between the second and third sub-bands. At finite bias voltage, the two-terminal conductance will only be quantized in integer values of $G_0$ if both $\mu_{\rm source}$ and $\mu_{\rm drain}$ occupy the same subband. This creates diamond shaped regions of constant conductance indicated by black dotted lines in Fig.~\ref{fig:fig4}a. At the tip of the diamond, the two dotted lines cross when $V_{\rm bias}$ is equal to the subband energy spacing $\Delta E_{sub}$ which can be estimated to be $\Delta E_{sub}\approx 16$ meV.

Phase-coherent transport is a basic requirement for the interferometric read-out schemes of the topological data, see Sec.~\ref{sec:perspective}. In order to study the transport properties of nanowire networks, the structures are transferred from the growth chip via a nanomanipulator in a scanning electron microscope (SEM) to the desired chip for device fabrication~\footnote{see movie at http://pubs.acs.org/doi/suppl/10.1021/acs.nanolett.7b00797}. An SEM image of the contacted nanowire hashtag structure is shown in the inset of Fig.~\ref{fig:fig4}b. The magnetoconductance of the device is shown in Fig.~\ref{fig:fig4}b where Aharonov-Bohm oscillations can be clearly seen. As expected, the period of the oscillations scales with the area of the hashtag devices demonstrating the phase coherent transport through the device. From temperature-dependent measurements, a phase coherence length of a few micrometers has been estimated.

As mentioned in the introduction, it is important to eliminate unintended subgap states in order to achieve a hard superconducting gap, which protects against thermal quasiparticle excitations. The density of states in the proximitized nanowire can be probed with tunneling spectroscopy by measuring the differential conductance as a function of source-drain voltage. A comparison between conventionally evaporated Al and epitaxially grown Al on InAs nanowires is shown in Fig.~\ref{fig:fig4}c. With epitaxially grown Al, the ratio of above-gap to subgap conductance is about 1/100. Interestingly, for thin Al on two facets, the superconducting gap remains hard, as shown in Fig.~\ref{fig:fig4}d. Furthermore, the size of the gap increases with decreasing Al thickness up to several nanometers~\cite{Court2008}. The devices with thin Al layer also withstand high magnetic fields (up to $\sim$ 3 T) as shown in Fig.~\ref{fig:fig4}d satisfying an important requirement for reaching the topological phase transition~\cite{Lutchyn10, Oreg10}.

Another important material requirement is a strong spin-orbit coupling (SOC) in the semiconductor. SOC is difficult to measure, especially in ballistic quasi-one-dimensional nanowires. A standard approach is via the quenching of weak antilocalization (WAL) by a small magnetic field. The field scale where the conductance peak at zero magnetic field associated with quantum-suppressed backscattering is quenched yields an estimate for the strength of SOC, within a rather complicated theoretical model~\cite{Iordanskii1994} that includes different scattering processes. Fig.~\ref{fig:fig4}e shows a fit to a WAL peak from a 2DEG, see Ref.~\cite{Shabani16} for details, from which the phase coherence length and Rashba SOC $\alpha$ can be extracted. The values measured in this InAs 2DEG are comparable to that measured in InAs and InSb nanowires~\cite{Shabani16, Weperen2015}. Similarly to the situation in nanowires, the density of states in 2DEGs can be measured via a tunnel probe, defined in the present case with a quantum point contact formed by top gates, see inset in  Fig.~\ref{fig:fig4}f. The gate-controlled constriction can be tuned from high transparency regime (i.e., a single conducting channel showing $4e^2/h$ two-terminal conductance) to the weak tunneling regime where the conductance measurement is proportional to the local density of states.

\section{Experimental signatures of Majoranas}

Even when all material requirements are satisfied, there remains a question of how MZMs can be identified experimentally, and how unambiguous are various signatures. One should appreciate that even when the system is in a trivial phase, without well-separated MZMs, fine-tuning of some system parameters can lead to experimental observations with signatures that resemble those expected for MZMs. A system in the topological phase, however,  shows these signatures in a robust way, i.e. they should persist under a moderate evolution of system parameters. It is this robustness that is the most characteristic trait of topological phases, and ultimately what makes these states useful for quantum information processing. In the following, we will discuss a {\em set} of experiments showing Majorana signatures, which increases our confidence that proximitized nanowires can be driven into a topological superconducting phase.

\begin{figure}
\includegraphics[width=\columnwidth]{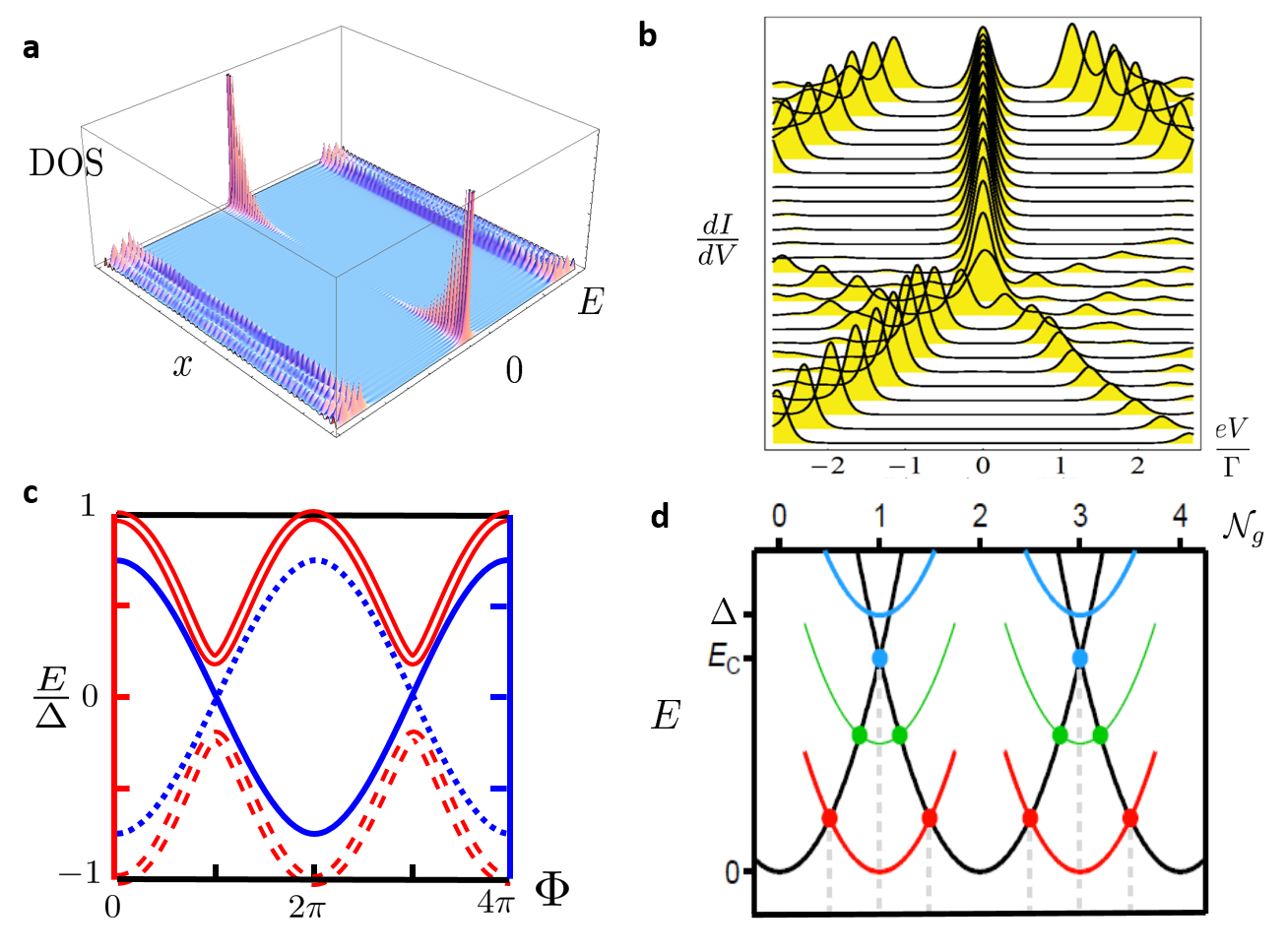}
\caption{ a) Simulated local density of states $\nu$ as a function of energy and coordinate along the nanowire for the proximitized nanowire in a topological phase with localized Majorana zero energy states at its ends. b) Differential tunneling conductance as a function of an in-plane magnetic field and applied voltage bias. Each line cut corresponds to a different magnetic field $B$. At the critical value of $B$-field, Majorana-induced zero-bias peak emerges. This feature is stuck to zero bias for a significant range of a magnetic field. Panels a) and b) adapted with permission from Ref.~\cite{Stanescu11}. c) The energy spectrum of Andreev bound states in SC-NW-SC junction as a function of SC phase difference $\Phi$ across the junction. Red and blue lines correspond to the energy spectrum for trivial ($B=0$) and topological ($B>B_c$) superconductors. In the trivial phase, there is an avoided level crossing at $\Phi=\pi, 3\pi$ due to doubly degenerate spectrum enforced by the time-reversal symmetry, whereas in the topological state this crossing is protected by particle-hole symmetry~\cite{kitaev01}. Dashed and solid blue energy states have different fermion parity. d) Charging energy of a proximitized nanowire as a function of the dimensionless gate voltage ${\cal N}_g$. Panel d) adapted with permission from Ref.~\cite{Albrecht16}. Blue and green lines correspond to the odd-charge states of a conventional (trivial) superconductor with $\Delta > E_C$ and $\Delta < E_C$, respectively. The red line represents the ground-state energy of an odd-charge state in a topological superconductor, in which case even- and odd-charge states are (almost) degenerate due to the presence of Majorana zero-energy modes. }
\label{fig:fig1}
\end{figure}

The simplest experiment involves tunneling conductance measurement of the local density of states at the nanowire ends~\cite{ZeroBiasAnomaly0, ZeroBiasAnomaly2,ZeroBiasAnomaly3,ZeroBiasAnomaly31, ZeroBiasAnomaly4, ZeroBiasAnomaly6, ZeroBiasAnomaly61, PradaPRB'12, RainisPRB'13}, see Fig.~\ref{fig:fig1}a. The dependence of the differential tunneling conductance on magnetic field is shown in Fig.~\ref{fig:fig1}b. As one can see, the characteristic zero-bias peak in tunneling conductance appears in a finite field (i.e. topological phase). The height of the zero-bias peak is predicted to be $2e^2/h$ at zero temperature~\cite{ZeroBiasAnomaly3, Fidkowski12, lutchyn_andreev13}. In addition to the zero-bias tunneling spectroscopy, a number of current correlations measurements probing the presence of MZMs has been proposed~\cite{Bolech2007, ZeroBiasAnomaly2, ZeroBiasAnomaly3, Golub2011, Haim2014, Haim2015, Liu2015a, Liu2015b}.

Fractional Josephson effect is another peculiar feature of MZMs. This effect corresponds to coherent tunneling of single electrons between two topological superconductors forming a Josephson junction. As a function of magnetic field, the spectrum of Andreev states should change from being $2\pi$-periodic in units of SC flux quantum to $4\pi$-periodic, see Fig.~\ref{fig:fig1}c. The former corresponds to tunneling of Cooper pairs whereas the latter appears due to the coherent single-electron tunneling processes. The characteristic crossing of two Andreev states at zero energy is protected by fermion parity in the junction~\cite{kitaev01}.  When parity is conserved, there is a true level crossing in the energy spectrum leading to a $4\pi$ periodicity of the supercurrent. In practice, finite-size effects lead to a small avoided level crossing, dynamical change of fermion parity may also occur due to unpaired electrons in the system. Therefore, supercurrent measurements should be performed at the sufficiently high frequency that these effects are avoided~\cite{kitaev01, Lutchyn10, Jiang'11, Pikulin2012, Houzet2013, PabloSanJose2013, Setiawan2017}.

Another signature of MZMs appears when semiconductor nanowire is proximitized by a floating superconductor (rather than grounded superconductor discussed so far) so that the overall system has a finite charging energy~\cite{Fu10, Zazunov2011, Hutzen2012, Hassler2015, Heck16, Lutchyn_charge'16, LutchynGlazman2017}. The corresponding Hamiltonian reads
\begin{align}
H_C=E_C (\hat {N}-{\cal N}_g)^2,
\end{align}
where $\hat {N}$ is the total number of electrons in the mesoscopic system (island), $E_C=e^2/2C_{\Sigma}$ with $e$ and $C_{\Sigma}$ being the electron charge and the total capacitance of the nanowire and its superconducting shell, ${\cal N}_g=C_g V_g/e$ is the dimensionless gate voltage with $C_g$ and $V_g$ being the gate capacitance and gate voltage, respectively.

In a conventional (s-wave) superconductor all electrons are paired in the ground state when the number of particles in a finite-size superconductor is even. Adding one more electron costs electrostatic energy as well as pairing energy $\Delta(B)$. Therefore, if $\Delta (B) > E_C$ the ground-state energy of the proximitized nanowire corresponds to an even-charge sector with $2e$-periodic dependence on ${\cal N}_g$, see Fig.~\ref{fig:fig1}d. The presence of a magnetic field leads to the suppression of the quasiparticle gap $\Delta(B)$, and the parity of the ground-state may have even- and odd-charge sectors, but the ground-state remains $2e$-periodic as long as $B < B_c$. In the topological phase ($B>B_c$), the periodicity of the ground-state energy changes due to the presence of MZMs and becomes $1e$-periodic~\cite{Fu10} as shown in Fig.~\ref{fig:fig1}d. The combination of this fact and elastic transport through a Coulomb-blockaded island (Majorana island) provides strong evidence for MZMs in the proximitized nanowires~\cite{Heck16}. Note that the corresponding transport through a metallic island (i.e. no MZMs) is dominated by inelastic processes~\cite{aleiner2002}.

An interesting scenario appears when Coulomb-blockaded island contains several pairs of MZMs. When such an island is coupled to several normal leads, Majorana-induced ground-state degeneracy leads to an exotic multi-channel Kondo effect~\cite{Beri2012, Altland2013, Michaeli2016}.

We emphasize again that all these properties characterize a topological phase and should persist in a finite parameter range. It may be possible to explain each of these physical observations separately within certain non-MZM scenarios by fine-tuning the system parameters. However, as discussed below, a large body of experimental data accumulated in different physical systems
allows one to rule out many of the false-positive interpretations in favor of a MZM scenario.

\begin{figure}
\center
\includegraphics[width=\columnwidth]{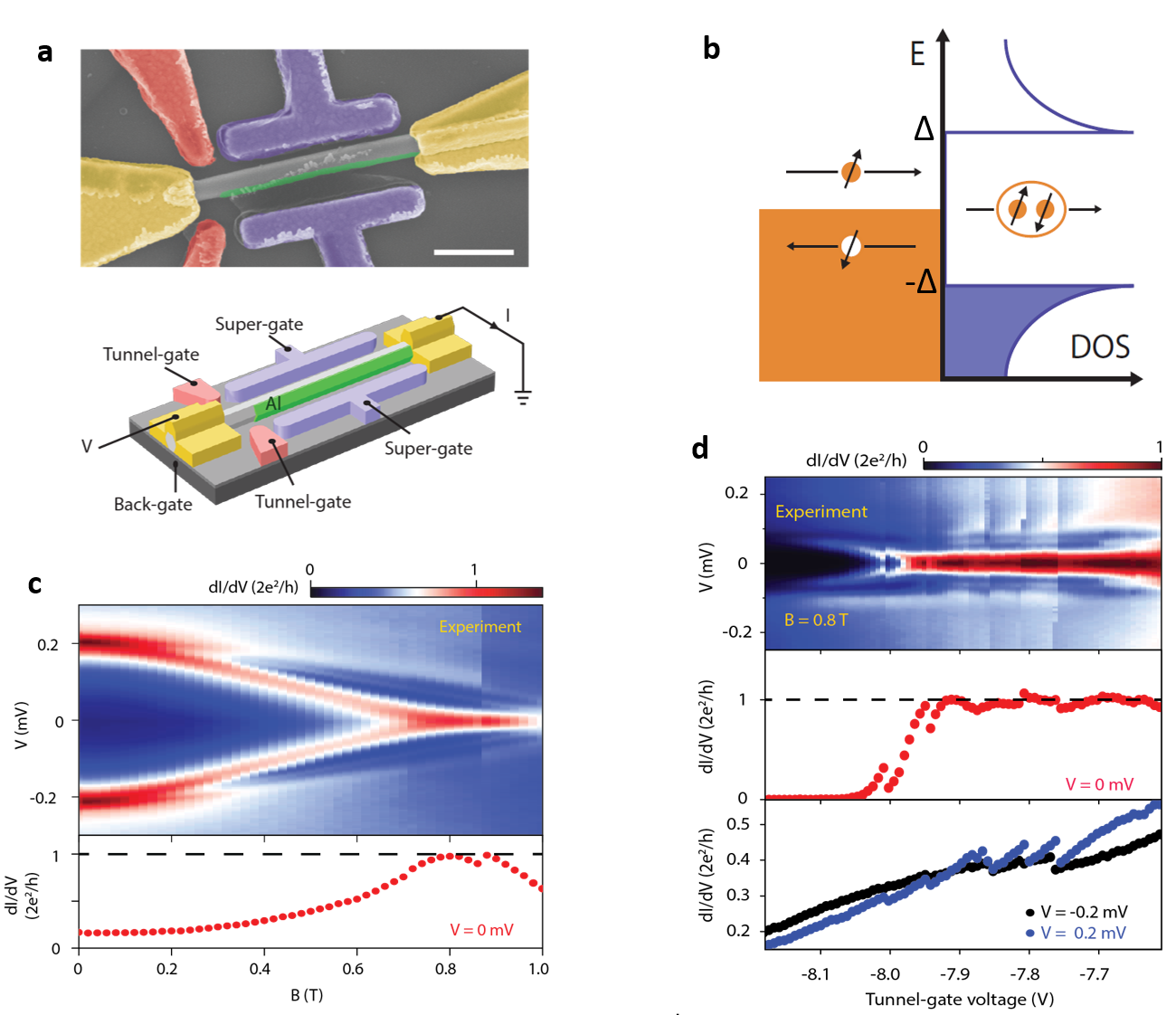}
\caption{Tunneling conductance in InSb/Al hybrid structures. a) The layout of the device: upper panel -  false-color electron micrograph, lower panel - device schematics. Here the side gates and the contacts are Cr/Au (10 nm/100 nm) and the Al shell thickness is ~10 nm. The substrate is p-doped Si which acts as a global back-gate. The barrier between the normal lead and the proximitized nanowire can be controlled by the two tunnel gate voltages. The scale bar is 500 nm. b) Schematic plot of Andreev reflection processes at a normal metal/superconductor interface. c) Tunneling conductance as a function of an in-plane magnetic field (up to 1 Tesla). The induced gap in this device is $\Delta_{\rm ind}= 0.2$meV; the zero-bias peak is visible between 0.7 and 1Tesla. Lower panel: zero-bias conductance line-cut as a function of magnetic field. Here the temperature is $T=20mK$. d) Dependence of the tunneling conductance on the tunnel-gate voltage at a fixed $B = 0.8$T. Lower panels show horizontal line-cuts at zero and finite bias. The zero-bias conductance is robust in large parameter regime and remains close to the quantized value of $2e^2/h$. Panels a), c) and d) adapted with permission from Ref.~\cite{Zhang2017}. }
\label{fig:fig5}
\end{figure}

\subsection{Zero-bias tunneling conductance measurements}

Tunneling spectroscopy is a conceptually simple method to measure local density of states and is well-suited to detect reconstruction of the energy spectrum across a topological phase transition. Given that MZMs are localized at the ends of the nanowire, the differential conductance measurement should exhibit robust zero-bias tunneling conductance peak in the topological phase. Typical device layout for tunneling spectroscopy measurement is shown in Fig.~\ref{fig:fig5}a. Low-bias transport in the normal-metal/superconductor junction is predominantly determined by the Andreev processes, during which an incident electron is reflected as a hole and a Cooper pair is added to the superconducting condensate, see Fig.~\ref{fig:fig5}b. In the non-topological phase ($B<B_c$) such processes are suppressed in a weakly-transparent junction. In contrast, the transmission probability becomes close to one in the topological phase~\cite{ZeroBiasAnomaly3, ZeroBiasAnomaly4, ZeroBiasAnomaly6, ZeroBiasAnomaly31, ZeroBiasAnomaly61, ZeroBiasAnomaly7, PradaPRB'12, RainisPRB'13, Liu2017}. The presence of MZMs leads to a {\it resonant} Andreev reflection, and, as a result, differential tunneling conductance at low temperatures (compared to tunnel broadening) acquires a universal value of $2e^2/h$~\cite{Law09, Fidkowski12, lutchyn_andreev13}.

Shortly after the theoretical proposals for proximitized nanowires~\cite{Lutchyn10, Oreg10}, the emergence of the zero-bias peak in a finite magnetic field was observed in Ref.~\cite{Mourik12}. Similar results were published in several subsequent studies~\cite{Deng12,Das12, Finck12, Churchill13, Chen2016} in InSb and InAs nanowires. First-generation experiments~\cite{Mourik12,Deng12,Das12,Finck12,Churchill13} reported the appearance of the zero-bias peak together with the closing of the bulk gap. The zero-bias feature persisted over a substantial range of magnetic fields and gate voltages and disappeared with the suppression of the superconductivity. Furthermore, the zero-bias peak disappeared when the in-plane magnetic field was aligned transverse to the wire axis, indicating that the effect depends on the interplay of Zeeman and spin-orbit couplings. All these observations are consistent with the Majorana scenario. At the same time, first-generation experiments revealed a large subgap density of states due to the interface disorder which co-existed with the zero-bias feature. It is challenging to separate the two contributions and various alternative scenarios have been proposed~\cite{Law2012, Bagrets2012, Pikulin2012, Kondo_Aguado, Lee_arxiv2013}.

Recent experiments with improved InSb/NbTiN~\cite{Zhang16, Chen2016} deposited interfaces, epitaxial InAs/Al \cite{Deng2016} nanowires and 2D InA/Al heterostructures \cite{Suominen2017, Nichele2017}, and epitaxial InAsSb/Al \cite{Sestoft2017} and InSb/Al~\cite{Zhang2017} interfaces have a much lower subgap density of states, see Fig.~\ref{fig:fig5}c, and exhibit ballistic transport properties allowing one to rule out the scenario of the disorder-induced zero-bias conductance peak. It is the material science progress reviewed in this work that enabled significant improvement in electric transport and allowed one to investigate subtle transport properties such as quantization of the zero-bias tunneling conductance. Indeed, previous experiments~\cite{Mourik12,Deng12,Das12,Finck12,Churchill13} reported zero-bias peak much smaller than the predicted value $2e^2/h$. Recent improvements of the samples quality led to the increase of the zero-bias conductance peak, and $G$ of order $e^2/h$ can be observed in InAs/Al~\cite{Vaitiekenas2017} and InSb/Al~\cite{Zhang2017} devices. Using improved InSb/NbTiN samples Chen {\it et al.}~\cite{Chen2016} mapped out the topological phase diagram through the dependence of zero-bias peak on the chemical potential and magnetic field. These measurements allow one to rule out many false-positive scenarios. However, it was pointed out in Ref.~\cite{Liu2017} that Andreev bound states may exhibit similar signatures. In order to distinguish between local Andreev bound states and delocalized MZMs, one has to carry out a detailed analysis of how the observed signal depends on temperature, tunnel coupling~\cite{Nichele2017, Setiawan2017}, which was reported experimentally recently in 1D and 2D heterostructure devices~\cite{Nichele2017, Zhang2017}, or consider more complex geometries, for instance with a quantum dot coupled to the end of a nanowire~\cite{Leijnse2011, Liu2011, Lee2013, Kondo_Aguado, Lee_arxiv2013, Cheng2014a, Clarke2017, Prada2017}, which allows a measurement of the non-locality of MZMs~\cite{Deng2017}. The authors of Ref.~\cite{Zhang2017} showed that tunneling conductance due to local Andreev states depends on the tunnel barrier height, see Fig.~\ref{fig:fig5}a, whereas Majorana-induced zero-bias peak is robust with respect to the tunnel coupling and, in fact, exhibits close to $2e^2/h$ conductance plateau~\cite{Zhang2017}.

Taken altogether, recent tunneling spectroscopy measurements provide a strong evidence for MZMs in semiconductor-superconductor heterostructures. The high quality of materials, as well as improved control in devices, open a possibility to study more exotic properties of MZMs.

\subsection{Coulomb blockade experiments with proximitized nanowires}

\begin{figure}
\center
\includegraphics[width=\columnwidth]{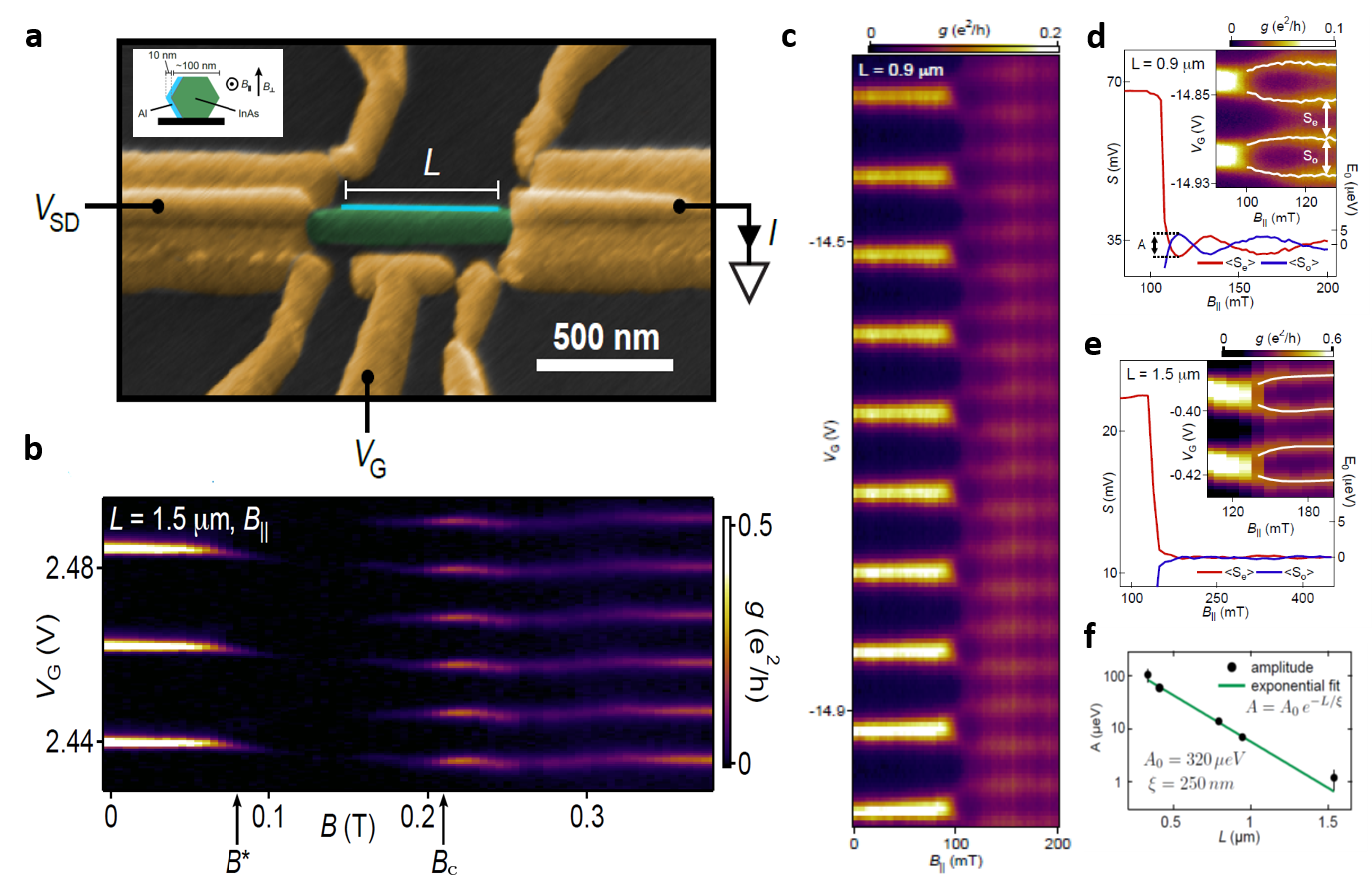}
\caption{Coulomb blockade experiment with proximitized nanowires. a) Electron micrograph (false color) of a device that is lithographically similar to the measured devices. Gold contacts (yellow), InAs nanowire (green), and two-facet Al shell of length $L$ (light blue). Applied voltage bias, $V_{SD}$, source-drain current, $I$, and gate voltage, $V_G$, are indicated. Inset: Cross section of a hexagonal InAs nanowire, showing orientation of Al shell and field directions $B_{||}$ and $B_{\perp}$. b) Two-terminal conductance, $g$, as a function of an external magnetic field $B_{||}$ and gate voltage for $L=1.5\mu$m. Coulomb peaks become dim at field $B^*$ and brighten at field $B_c$. The field $B^*$ is the characteristic scale for the crossover from coherent Cooper-pair to incoherent single-electron tunneling regime whereas the field $B_c$ corresponds to the topological phase transition. At $B>B_c$ transport is dominated by the coherent single-electron tunneling processes due to MZMs~\cite{Fu10, Heck16}. c) Zero-bias conductance, $g$, as a function of gate voltage, $V_G$, and parallel magnetic field, $B_{||}$, for $L=0.9\mu$m device, showing a series of $2e$-periodic
Coulomb peaks below $100$mT and $1e$ nearly-periodic peaks above
100 mT. d) and e) Average peak spacing for even and odd Coulomb valleys, $S_{e/o}$, as a function of magnetic field for L = 0.9 $\mu$m and L = 1.5 $ \mu$m, respectively. Inset: Even and odd peak spacings, $S_{e/o}$, are indicated by arrows.
f) Oscillatory amplitude, $A$, plotted against the proximitized nanowire length, $L$, for 5 devices from 330 nm to 1.5 $\mu$m (black dots) and exponential fit to $A = A_0 \exp(-L/\xi)$ with
$A_0$= 300 $\mu$eV and $\xi = 260$ nm. Error bars indicate uncertainties propagated from lever arm measurements and fits to peak maxima. Panels a) - f) adapted with permission from Ref.~\cite{Albrecht16}.}
\label{fig:fig6}
\end{figure}

The significant improvement of the interface quality in epitaxially-grown proximitized nanowires opens a possibility to study an interplay of topological superconductivity and Coulomb blockade phenomena~\cite{Higginbotham15, Albrecht16}. The device consists of a mesoscopic proximitized nanowire with a finite charging energy $E_C$ and is coupled to two normal leads, see Fig.~\ref{fig:fig6}a. In the Coulomb blockade regime, the charge on the island can be controlled by changing the gate voltage $V_g$, as well as electrostatically tuning single-channel barriers. Both these knobs enable one to investigate the crossover from strong to weak Coulomb blockade regimes and study quantum charge fluctuations in the island. Quantum charge fluctuations allow one to identify the topological phase since they have an imprint of the physical state of the system~\cite{Lutchyn_charge'16}.

As explained above, physical observables change the dependence on the dimensionless gate voltage ${\cal N}_g$ across the topological phase transition~\cite{Fu10} - charge periodicity changes from $2e$ to $1e$. Therefore, two-terminal conductance $G({\cal N}_g)$ through the island at zero-bias should exhibit the crossover between these two limits as a function of magnetic field. Theory~\cite{Heck16} predicts that the conductance   is determined by different processes as magnetic field is increased from zero: (I) Cooper-pair dominated regime for $B < B^*$; (II) sequential tunneling regime ($B^* < B < B_c$); (III) Majorana-dominated regime for $B>B_c$.  Here $B^*$ is the field at which ground-state can have different parity, i.e. $\Delta(B^*)=E_C$. In the first limit, the ground state of the system involves only even-parity sectors. Therefore, the zero-bias conductance peaks appear at the degeneracy points between nearest even-charge states, see  Fig.~\ref{fig:fig6}b. In the case (II), ground state as a function of ${\cal N}_g$ changes its parity between even- and odd-charge sectors. Zero-bias conductance peaks now correspond to the change in ground-state parity. As a result, each conductance peak splits into two at $B=B^*$, see Fig.~\ref{fig:fig6}b, and this splitting is increasing with magnetic field until it reaches $B=B_c$. Note that at these degeneracy points the dominant conduction mechanism is the resonant transfer of single electrons rather than Cooper pairs. In this intermediate regime, the peak conductance at low temperatures becomes suppressed, see Fig.~\ref{fig:fig6}b, and, in fact, should vanish in the thermodynamic limit $L \rightarrow \infty$~\cite{Averin'93, Heck16}. Finally, in the topological regime (III) peak positions occur at half-integer values of ${\cal N}_g$, independent of the magnetic field (up to finite-size corrections). In this limit, the dominant contribution to the peak conductance originates from the resonant tunneling via MZMs. Due to the non-local nature of the fermionic mode shared between two MZMs, the height of conductance peak is much larger than in the case (II) and is independent of $L$.

In order to investigate the non-local nature of the topological ground-state degeneracy, one may study the splitting energy originating from an exponentially-small Majorana wavefunction overlap: $\delta E \propto \exp(-L/\xi)$. This difference for even- and odd-parity states results in a small shift of the charge degeneracy points. Thus, the peak spacing within the Majorana scenario in the even- ($S_e$) and odd-charge ($S_o$) sectors should be different, whereas in the normal-island case $A=S_e-S_o$ is negligibly small. As shown in Figs.~\ref{fig:fig6}c-f,  odd-even spacing $A$ does depend on the nanowire length and, thus, the experimental results~\cite{Albrecht16} are consistent with the MZM scenario. The splitting measurement provides important information about the physical system (e.g., superconducting coherence length), and it would be quite interesting to analyze this issue more systematically in the future and to compare experimental results with a microscopic simulation of the proximitized nanowires~\cite{Chiu2017}. \\

\section{Perspective for topological quantum computation}\label{sec:perspective}

During the last five years we witnessed the birth of a new field of research - mesoscopic topological superconductivity~\cite{Fu10, Alicea11, Zazunov2011, Hutzen2012, Hyart13, Beri2012, Altland2013, Altaland2014, Cheng2014, Dong2015, Albrecht16,Landau16, Hoffman2016, Michaeli2016, Heck16, Aasen16, Plugge16a, Vijay2016, Plugge16b, Karzig16, Lutchyn_charge'16, LutchynGlazman2017}, and saw an impressive growth of experimental explorations of various superconducting heterostructures~\cite{Mourik12,Deng12,Das12, Finck12, Churchill13, Chang14, Krogstrup15, Higginbotham15, Kammhuber16, Shabani16,  Zhang16, Deng2016, Albrecht2017, Gazibegovic2017}. The interplay of mesoscopic and topological physics enables one to manipulate quantum information stored in topological (non-local) degrees of freedom in a way that minimizes decoherence effects. Remarkable experimental progress, rapid improvement in the materials quality as well as our enhanced theoretical understanding of the microscopic details are encouraging developments for the prospect of building topological qubits in superconductor-semiconductor heterostructures.

\begin{figure}
\center
\includegraphics[width=\columnwidth]{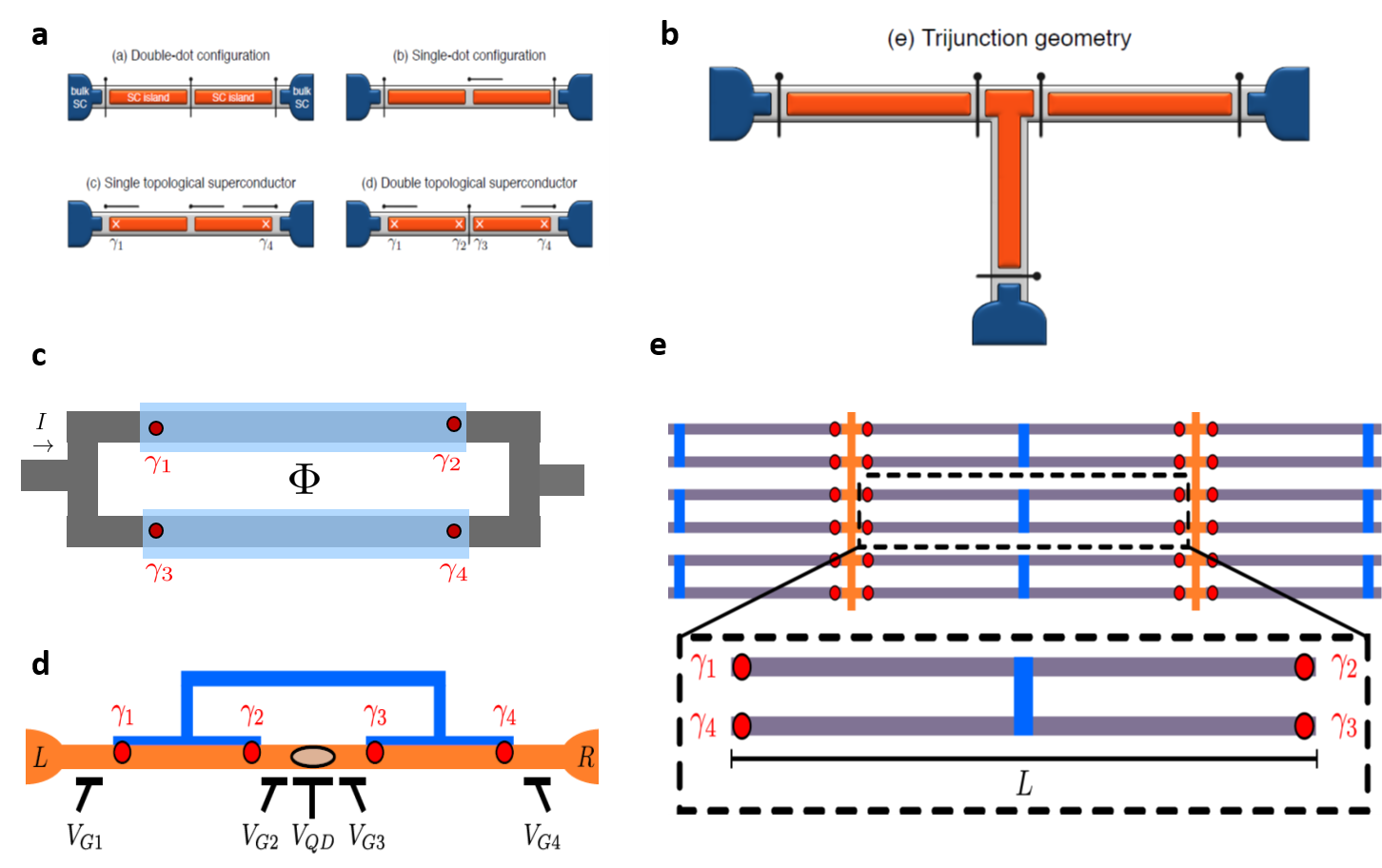}
\caption{Proposed future experiments probing key properties of topological superconductors for quantum computing applications. a) Majorana fusion experiment. The device, consisting of two mesoscopic Majorana islands (orange) with gate-tunable tunnel barriers (valves). Charging energy of the island can be quenched by lowering the barrier to a bulk superconductor (dark blue). The middle barrier connects the islands. When the middle valve is completely open, one effectively achieves single-island configuration. The specific protocol of opening (closing) the valves allows one to investigate Majorana fusion rules. b) Topological T-junction for braiding experiment. Panels a) - b) reproduced with permission from Ref.~\cite{Aasen16}. Anyonic state teleportation, equivalent to physically braiding Majoranas, can be achieved by closing (opening) the valves as explained in Ref.~\cite{Alicea11}. c) Majorana interferometer - a basic element of the Majorana surface code~\cite{Landau16, Vijay2016, Plugge16a}. Coherent transport through the interferometer probes non-local nature of Majorana zero modes. The position of MZMs $\gamma_i$ is represented as red dots. d) Minimal setup for a quasiparticle-poisoning-protected topological qubit~\cite{Plugge16b, Karzig16}. Mesoscopic superconducting island hosting four MZMs (red dots) is in a Coulomb blockade regime with fixed total charge. The island has two-fold ground-state degeneracy and represents the simplest topological qubit. Semiconductor quantum dot, coupled to the qubit via gate-tunable barriers (i.e. the barriers can be controlled with gate voltages $V_{Gi}$ ), is used to perform projective measurements. Parity-dependent charge fluctuations between SM dot and SC island shift energy level in QD, allowing one to distinguish between different degenerate states of the qubit. The energy difference between even- and odd-charge states can be tuned with $V_{QD}$.  e) Scalable architecture for Majorana-based quantum computing. Topological qubits are coupled together via semiconducting quantum dots. Quantum computation is performed using measurement-only protocol, which is facilitated by tunable couplings between Majorana zero modes and nearby semiconductor quantum dots~\cite{Bonderson08b}. Panel e) reproduced with permission from Ref.~\cite{Karzig16}. }
\label{fig:fig7}
\end{figure}

Quantum gates within the topological quantum computation approach rely on non-local transformations within the degenerate ground-state manifold, which can be performed either by adiabatically changing physical parameters of the system~\cite{Alicea11, Hyart13, Bonderson2013, Barkeshli2015, Aasen16}, or using the projective measurements~\cite{Bonderson08b, Bonderson2009, Plugge16b, Karzig16}. We now outline several immediate directions, which should be explored both experimentally and theoretically in order to test the necessary ingredients for a successful operation of a futuristic quantum computer, see Fig.~\ref{fig:fig7}.

As a first step, we need to establish that the exotic non-local properties of MZMs are indeed realized in the experimental system. A hallmark of topological states of matter is ground-state degeneracy which so far has not been tested experimentally.  Fusion and braiding rules associated with zero modes define a topological phase and are of fundamental importance for topological quantum computing. The fusion and braiding experiments can be performed in a small network of nanowires as shown in  Fig.~\ref{fig:fig7}a and b. The fusion experiment is simpler than braiding as it does not require an exchange of MZM position but nevertheless reveals topological properties of the system.

In order to test fusion rules, we follow the ideas outlined in Refs.~\cite{Alicea11, Aasen16} and consider a device consisting of two Majorana islands connected together and to the superconducting leads by gate-tunable tunnel barriers (valves), see Fig.~\ref{fig:fig7}a. The effective charging energy of the left (right) island can be reduced by lowering the left (right) barrier to a  superconducting lead. Let us denote open (low barrier) and closed (high barrier) valve as ``1'' and ``0'', respectively. The string ‘010’, for example, represents the configuration where the left and right valves are closed whereas the middle valve is open. This configuration of valves effectively merges the left and right islands to a single island disconnected from the superconducting reservoirs.
As shown in Ref.~\cite{Aasen16}, two distinctive sequences of open and closed valves configurations, differing only by one intermediate configuration, should eventually lead to different final charge configurations on the islands. The latter reflects the fusion rules of non-Abelian MZMs. While the sequence $a_1$: $010 \rightarrow 000 \rightarrow 101 \rightarrow 111$ leads to a definite deterministic charge configuration on the left and right islands; the sequence $a_2$: $010\rightarrow 111 \rightarrow 101 \rightarrow 111$ results in an equal superposition of two distinct charge distributions. A measurement of the charge on the island following an execution of sequence $a_1$ should be deterministic. In contrast, after the execution of sequence $a_2$ one would expect to find two charge configurations with equal probability.

In order to perform braiding operations three Majorana wires forming a T-junction, see Fig.~\ref{fig:fig7}b,
are necessary~\cite{Alicea11, Aasen16}. The elementary sequence requires controlling five valves, i.e. keeping the leftmost and the rightmost valves open and tuning the remaining three valves~\cite{Aasen16}. Following the aforementioned notations, the sequence $b_1$ - $00000 \rightarrow 01100 \rightarrow 00110 \rightarrow 00000$ - results in braiding of the two MZMs next to the operative left and right valves. Here the numbering of valves goes from left to right with the middle number representing the coupling of the middle island to the lead. Note that double braiding operation should always lead to a flip of the parity of the qubit which is another non-trivial test.

At the heart of measurement-only topological quantum computation is a projective measurement of the joint fermion parity encoded in a pair of MZMs. In this scheme, an effective braiding between two non-Abelian anyons is achieved by measuring the topological charge of anyonic pairs (i.e. fermion parity) rather than physically exchanging their position~\cite{Bonderson08b, Bonderson2009}. Interferometric measurement is a promising way to distinguish between even and odd fermion parities, see, for example, the device shown in Fig.~\ref{fig:fig7}c. As explained above, transport through a proximitized nanowire in a Coulomb blockade regime changes its periodicity with the gate charge from $2e$ and $1e$ as a function of applied magnetic field. In the topological regime~(large $B$-field), charge transport is dominated by the coherent single-electron transmission through MZMs~\cite{Fu10, Heck16, Lutchyn'17} which is very different from the inelastic transport in a normal-metal island although both systems exhibit $1e$ periodic signal. Thus, the interferometric measurement should be able to distinguish between different tunneling processes: transmission via MZMs or tunneling through some other delocalized subgap states. The former preserves electron coherence and, therefore, should lead to the dependence on the enclosed magnetic flux $\Phi$ whereas the latter does not. Although this setup cannot access properties of topological qubits (since the MZM parities are fixed by the charging energy), such an experiment is a crucial test of the concept of measuring MZM states using coherent links and represents a basic building block of Majorana-based surface~\cite{Landau16, Vijay2016, Plugge16a} and color~\cite{Litinski2017} codes.

In order to perform quantum information processing the mesoscopic superconducting island should support at least four MZMs maintaining ground-state degeneracy with a fixed number of electrons due to a large charging energy. The simplest topological qubit is depicted in Fig.~\ref{fig:fig7}d. The projective measurement can be performed by coupling MZMs to a central semiconductor quantum dot, see Fig.~\ref{fig:fig7}d, and measuring, for example, the quantum capacitance of the system. Indeed, as shown in Refs.~\cite{Flensberg2011a, Plugge16b, Karzig16}, virtual exchange of electrons between the semiconductor quantum dot and Majorana island results in a parity-dependent contribution to the ground-state energy which ultimately leads to a change of QD charge distribution. Majorana islands supporting six MZMs open the possibility to perform a measurement-only braiding operation by the sequential measurement of the fermion parity of different pairs of MZMs~\cite{Karzig16}.

Finally, an array of topological qubits coupled together via semiconducting quantum dots represents a scalable architecture for Majorana-based quantum computing~\cite{Karzig16}. An example of such architecture is shown in Fig.~\ref{fig:fig7}e.  Quantum computation is performed using the measurement-only protocol which is facilitated by tunable couplings between MZMs and nearby semiconductor quantum dots. The detailed protocol for implementation of Clifford gates using a limited set of projective measurements in this architecture is discussed in Ref.~\cite{Karzig16}. Combining the topologically-protected Clifford gates with the ability to produce and distill magic states~\cite{Bravyi'05, Bravyi'06, SauWireNetwork, JiangKanePreskill, TopologicalQuantumBus,  Karzig'15, Clarke2016} should ultimately lead to universal quantum computation with MZMs.

\section{Summary}

The search for Majorana zero modes in superconductor-semiconductor heterostructures is evolving at a rapid pace and now encompasses a broad range of disciplines: from material science to condensed matter physics and quantum information science. In less than a decade since the theoretical proposals~\cite{Lutchyn10, Oreg10}, we have accumulated a compelling evidence for MZMs in proximitized nanowires. Most of the experiments  searching for Majoranas focused on tunneling spectroscopy. The recent work on InSb/Al proximitized nanowires~\cite{Zhang2017} reported robust $2e^2/h$ conductance consistent with the Majorana scenario. Chen et al.~\cite{Chen2016} mapped out the phase diagram through the dependence of zero-bias peak on chemical potential and magnetic field. The experiments with floating proximitized nanowires~\cite{Albrecht16} set the stage for the exploration of Coulomb blockade effect in mesoscopic topological superconductors. A measurement of $4\pi$ fractional Josephson effect was recently reported in Ref.~\cite{Geresdi2017}. While these experiments strongly support the existence of the emergent MZMs in proximitized nanowires, a direct observation of the non-Abelian properties associated with Majoranas is still lacking. In Sec.~\ref{sec:perspective}, we outlined future experiments for the detection of more exotic properties of MZMs such as braiding and fusion rules. Observing two MZMs at the ends of a single wire simultaneously is a necessary condition for that.

Experiments with proximitized nanowires do not only contribute to our understanding of topological phases of matter but also have significant theoretical and practical ramifications that span beyond condensed matter physics. The next-generation experiments with Majorana nanowires will set the stage for the manipulation of quantum information in topological systems. Scalable designs for topological quantum computation based on superconducting islands hosting multiple pairs of MZMs have been recently put forward~\cite{Plugge16b, Karzig16}.

Experimental demonstration of the exotic physics associated with MZMs, such as non-Abelian statistics, would constitute a breakthrough for all of fundamental physics, as well as pave the way for the next important milestone - a validation of a topological qubit operating in the exponentially-protected regime and exhibiting long coherence times. These accomplishments would lead to the exploration of different ideas for topological quantum memory and fault-tolerant universal quantum computation.


\appendix

\section{Box 1: A minimal model for 1D topological superconductor}\label{box}

The minimal model for 1D topological superconductor involves semiconductor nanowire with strong spin-orbit coupling proximity-coupled to a conventional (s-wave) superconductor~\cite{Lutchyn10, Oreg10}. The effective low-energy Hamiltonian for such a system is given by
\begin{align}
H&=H_{\rm SM}+H_{\rm P},\label{eq:H0}\\
\!\!H_{\rm SM}\!&=\sum_{\sigma, \sigma'}\!\int_{0}^{L} \!\!\!d x \psi_{\sigma}^\dag(x)\!\left(\!-\!\frac{\hbar^2\partial_x^2}{2m^*}\!-\!\mu\!+\!i \hbar \alpha \hat{\sigma}_y \partial_x\!+\!V_Z \hat{\sigma}_x\!\right)_{\sigma\sigma'}\!\!\!\!\!\psi_{\sigma'}(x),\nonumber\\
H_{\rm P}&= \int_{0}^{L} \!d x \left(\Delta_{\rm ind} \psi_{\uparrow}^\dag(x) \psi^\dag_{\downarrow}(x)+h.c. \right).\nonumber
\end{align}
Here $\hbar$ is the Planck constant; $m^*$, $\mu$ and $\alpha$ are the effective mass, chemical potential, and Rashba spin-orbit coupling in the semiconductor nanowire of length $L$, respectively. $V_Z$ is the Zeeman splitting due to the applied magnetic field $B_x$:  $V_Z\!=\!g \mu_B B_x$ with $g$ and $\mu_B$ being the Lande g-factor and Bohr magneton. $B_x$ is an external magnetic field applied along the nanowire, and $\hat{\sigma}_i$ are Pauli matrices. The proximity to the s-wave superconductor is effectively described by the Hamiltonian $H_P$ with $\Delta_{\rm ind}$ being the induced pairing gap. Typical material parameters for InAs and InSb semiconductors are provided in Table \ref{Table1} whereas values for the induced gap are given in Table \ref{Table2}.

Although semiconductor nanowire is coupled to an s-wave superconductor, the presence of the spin-orbit coupling leads to spin-momentum locking and results in a mixed Cooper-pair wave function consisting of the singlet and triplet pairings~\cite{GorkovRashba2001}. When Zeeman term opens a large gap in the spectrum (i.e. $V_Z > \sqrt{\mu^2+\Delta_0^2}$), the singlet pairing component is suppressed, and the superconducting state forming at the interface has p-wave symmetry of the order parameter. This state is adiabatically connected to Kitaev's model~\cite{kitaev01} and supports MZMs at the opposite ends of the wire.


%

\end{document}